\documentclass{PoS}

\newcommand\units{\,\mathrm}

\def\eVdist{\kern-0.06667em}

\title{Particle physics perspective}

\ShortTitle{Particle physics perspective}

\author{\speaker{Halina Abramowicz}\thanks{Partly supported by the Israel Science Foundation and the I-CORE Program No. 1937/12.}\\
        Tel Aviv University\\
        E-mail: \email{halina@post.tau.ac.il}}

\abstract{A personal perspective on the future challenges of research in particle physics is presented with emphasis on the role of DIS physics in this endeavour.}

\FullConference{XXIV International Workshop on Deep-Inelastic Scattering and Related Subjects\\
		11-15 April, 2016\\
		DESY Hamburg, Germany}

\begin{document}

\section{Legacy of HERA}

\begin{figure}
\centering
\includegraphics[width = 0.6\linewidth]{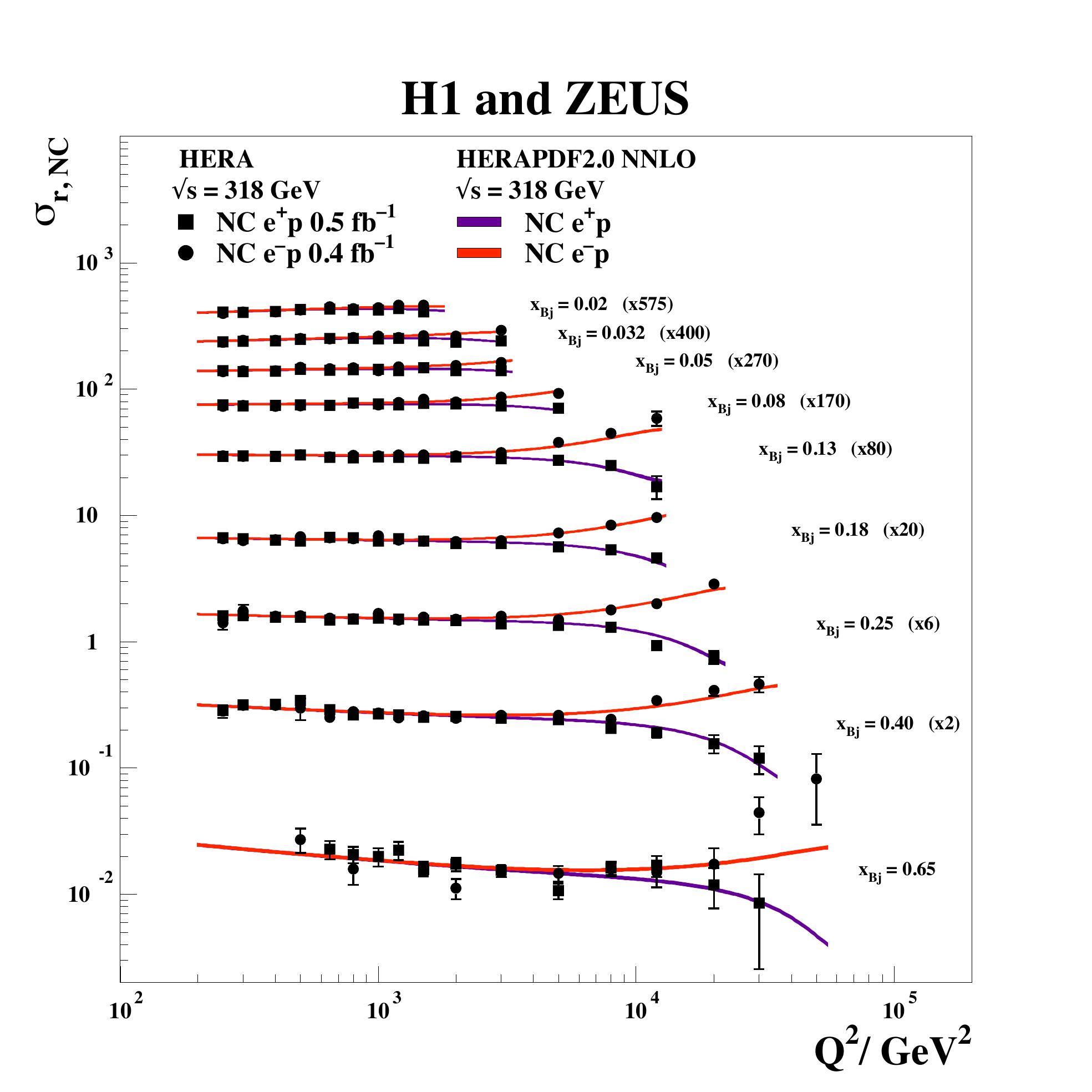}
\caption{Reduced neutral current $e^+p$ and $e^-p$ cross section as a function of $Q^2$ at fixed values of $x$ indicated in the plot, compared to NLO QCD evolution fits to the same data. The measurements are the results of combining H1 and ZEUS results.}
 \label{fig:F2}
\end{figure}

\begin{figure}
\centering
\includegraphics[width = 0.5\linewidth]{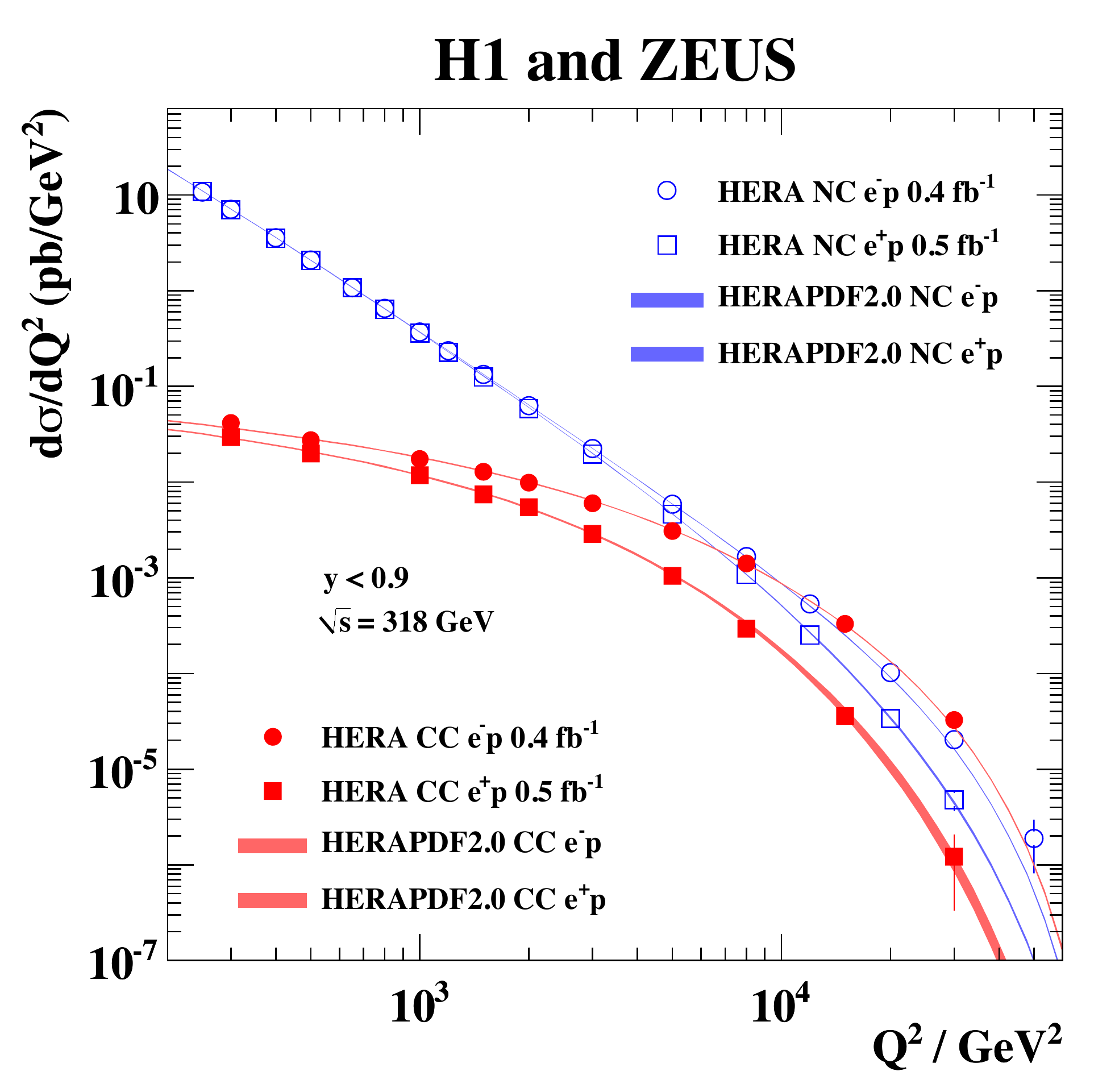}
\caption{Integrated charged current and neutral current, $e^+p$ and $e^-p$, cross sections as a function of $Q^2$ compared to NLO QCD expectations obtained from evolution fits to the differential xross section measurements. The data are the results of combining H1 and ZEUS results. }
 \label{fig:sigmanc+cc}
\end{figure}

The DIS series of workshops on deep-inelastic scattering came into existence during the era of $ep$ interactions at HERA to guide the experimental community in extracting from HERA data the best possible measurements of the structure of the proton. On the way, we also learned a lot about the structure of strong interactions in the perturbative regime of QCD. We had moments of excitement, not unlike at LHC, when we found that the inclusive electromagnetic structure function of the proton, $F_2$ was rising with decreasing $x$~\cite{lowx-H1,lowx-ZEUS} and that the diffractive processes could contribute to inelastic scattering at the leading twist level~\cite{diffH1,diffZEUS}. Not unlike the LHC, we found border-line significant excesses of events~\cite{largex-ZEUS,largex-H1} which ultimately were not confirmed by increasing the statistical significance of the measurements. In spite of some hints that the DGLAP~\cite{DGLAP} dynamics could be augmented by processes expected in the BFKL~\cite{BFKL} regime~\cite{forward-jets}, all of which could only be established through approximate implementation in Monte Carlo codes, the community came out of HERA convinced that parton density functions (PDFs) in the proton could safely be extracted from measurements of $F_2$ in neutral (NC) and charged (CC) current interactions through the usual DGLAP evolution fits~\cite{HERAF2}. The legacy of HERA can thus be summarized in two figures, Fig.~\ref{fig:F2} where the combined results of the reduced inclusive NC cross section by H1 and ZEUS are presented as a function of $Q^2$ at fixed values of Bjorken-$x$ for $e^+p$ and $e^-p$ interactions. and in Fig.~\ref{fig:sigmanc+cc} where the NC and CC cross sections integrated over $x$ are presented as a function of $Q^2$. The latter figure is often interpreted as the proof of unification of electromagnetic and weak interactions as the CC cross section due to $W^\pm$ exchange becomes equal in size to the NC cross section dominated by $\gamma^\star$ exchange (with a small admixture of $\gamma^\star-Z$ interference terms). 

The above summary does not give justice to the extensive measurements that were conducted at HERA on photoproduction, properties of hadronic final states, exclusive processes, jets production. All these measurements confirmed the basic understanding of QCD at high energy in the perturbative regime. Some, like the observation and studies of exclusive vector meson production or deep virtual Compton scattering, a minute contribution to the total cross section, are now taken up by the fixed target program at CERN (COMPASS) and JLab to explore the three dimensional structure of the proton.

\section{Status of the Standard Model}

\begin{figure}
\centering
\includegraphics[width = 0.8\linewidth]{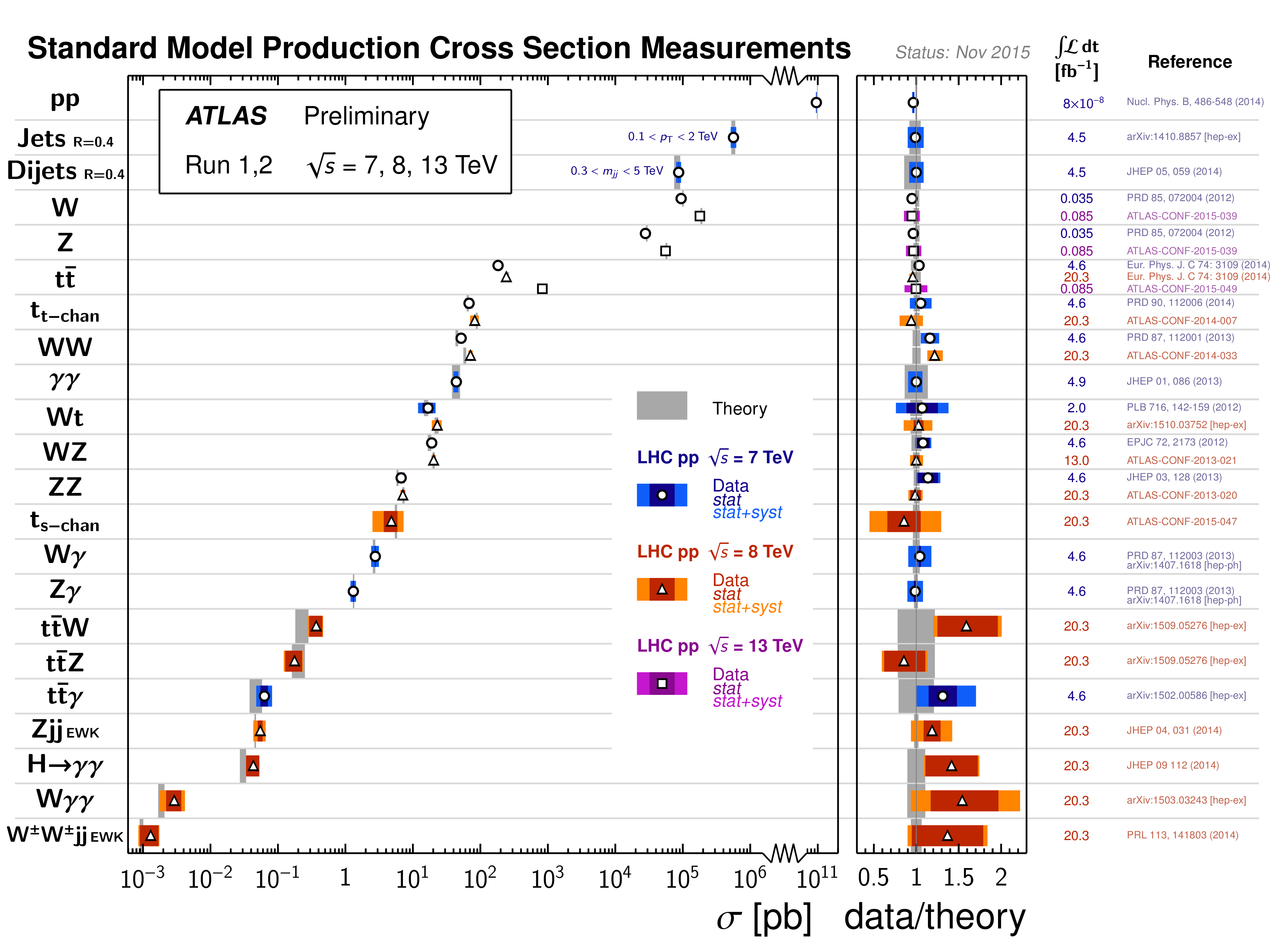}
\caption{Compilation of fiducial cross sections for various final states in $pp$ interactions at the LHC, as indicated in the figure, compared to expectations of the Standard Model, as compiled by the ATLAS experiment.}
 \label{fig:SMLHC}
\end{figure}

The HERA legacy constitutes an essential input in the validitation (or invalidation) of the Standard Model at the LHC. An amazing agreement between measurements and expectations is observed as summarized for example in Fig.~\ref{fig:SMLHC} over almost 14 orders of magnitude in cross section values for $pp$ interactions at center of mass energies $\sqrt{s}$ of 7, 8 and $13 \units{TeV}$. This is a highly non-trivial achievement, which requires input not only in the form of PDFs but also a good understanding of the production of final states, not to mention the harsh running conditions and the complexity of LHC detector systems. 

The complexity of the undertaking has been summarized by Pavel Nadolsky in his presentation at this meeting entitled ``Parton distributions from HERA to the LHC'' in his ``Maps of QCD concepts''~\cite{talk:nadolsky-p14}.

\begin{figure}[h]
\centering
\includegraphics[width = 0.5\linewidth]{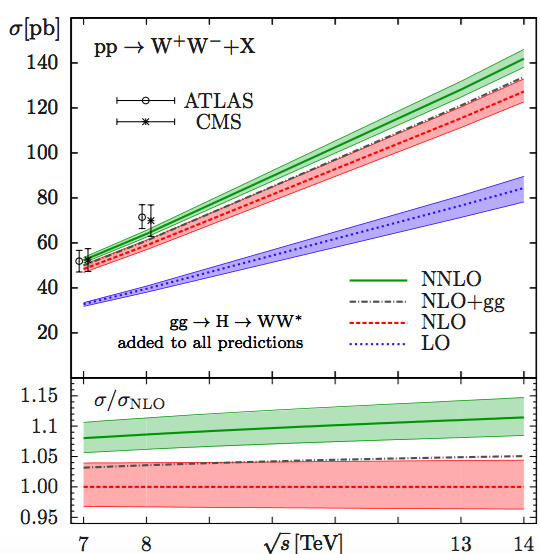}
\caption{The $W^+W^-$ production cross section in $pp$ interactions as a function of center of mass energy $\sqrt{s}$ compared to QCD expectations at various orders of perturbative expansion, as indicated in the plot.}
 \label{fig:WWexcess}
\end{figure}

The immense challenges have mobilized the theoretical community to push the precision level of QCD calculations. The NLO calculations, including NLO and parton shower (PS) matching, are readily available for most processes  and substantial progress has been achieved in NNLO calculations. Future directions are towards NLO+PS as a new standard for event generators, NNLO automation of calculations and even to go beyond NNLO as presented by Thomas Gehrmann
jn ``Advances in QCD predictions''~\cite{talk:gehrmann} at the opening of this workshop. The new standards explain the observed $W^+W^-$ production excess in the early LHC data compared to the SM expectations~\cite{WWexcess} as due to missing higher order corrections in the calculations~\cite{GehrmannWW} as shown in Fig.~\ref{fig:WWexcess}. The achieved progress is also reflected in the fact that today the estimated theoretical uncertainty on the Higgs production cross sections at the LHC energies has shrunk to 3\%~\cite{talk:huston-at-PDFLHC-april2015}.

There is progress in unravelling the spin content of the proton with new data coming from Brookhaven,  CERN and JLab as summarized by 
Ralf Seidel~\cite{talk:ralf-seidel}. Progress has also been made on the 3D structure of the nucleon.

\section{Glossary of future facilities}
The European community of particle physicists is readying for the European Strategy Update. It is already known that it will start after completion of the LHC RunII in 2018 with conclusions expected sometime in the mid 2020. The natural question is then what next? 

\begin{figure}[h]
\centering
\includegraphics[width = 0.7\linewidth]{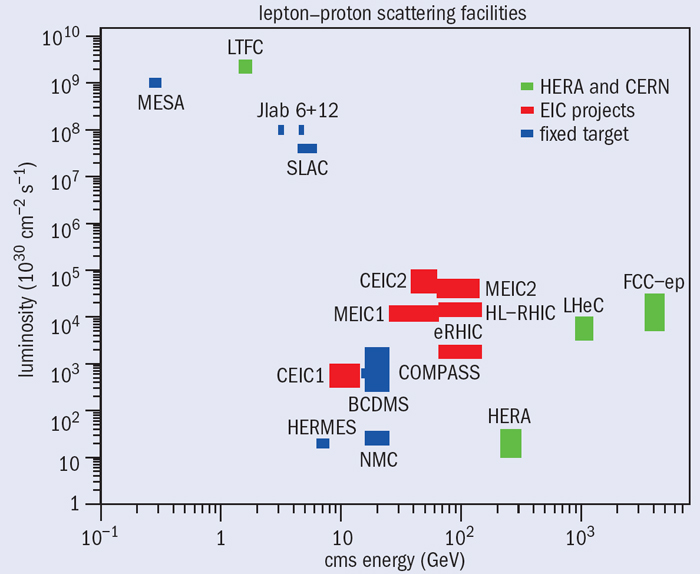}
\caption{Schematic representation of various $ep$ and $e$-ion facilities, completed or planned for the future, in the plane of luminosity and center of mass energy.}
 \label{fig:eAfacilities}
\end{figure}

As usual the choice is between a hadron machine, an $e^+e^-$ collider or an $ep$ collider. The following are the presently discussed facilities:
\begin{itemize}
\item Hadron colliders
    \begin{enumerate}
    \item HE-LHC - this would be the LHC with present magnets replaced  by high field magnets which are being developed for the Future Circular Collider (FCC). This could double the center of mass energy to $28 \units{TeV}$;
    \item FCC - this is to be a $pp$ circular accelerator located in an 80 to $100 \units{km}$ tunnel in and around Geneva based on 16 to 20$\units{T}$ magnets~\cite{FCC};
    \item SppC - a machine proposed by the Chinese community, similar to the FCC, which would be located in a $50 \units{km}$ tunnel and reach 50 to $100 \units{TeV}$~\cite{precdr-SppC};
    \item SSC - there is an option to build a $270 \units{km}$ tunnel at the location of the former SSC project with existing superconducting technology which would allow to achieve $100 \units{TeV}$ $pp$ interactions, to be upgraded to $300 \units{TeV}$ with FCC type magnets~\cite{1402.5973}.
   \end{enumerate}
\item $e^+e^-$ colliders
     \begin{enumerate}
      \item ILC - the International Linear Collider, technologically ready to be built~\cite{ILC-TDR}, is intended to be a $500 \units{GeV}$ center of mass energy collider~\cite{ILC-schedule} with a further possible upgrade to $1 \units{TeV}$ by extending its length. The project is presently considered to be hosted in Japan;
     \item CLIC - a linear collider with a two beam acceleration scheme and conventional magnets~\cite{CLIC-CDR}, would start at $380 \units{GeV}$ center of mass energy collisions and then be extended in length to reach $3 \units{TeV}$~\cite{CLIC-schedule};
     \item FCC-$ee$ - the FCC tunnel could be used to provide a circular collider with center of mass energies of $240$ to $350 \units{GeV}$ for a Higgs factory
     or  a GigaZ factory at lower energies~\cite{FCC-ee};
     \item CEPC - the Chinese version of the FCC-$ee$ in the same tunnel as the SppC with  center of mass energies of $240  \units{GeV}$;
     \item SSC - if resurrected the  $87 \units{km}$ circular tunnel of the former SSC could host a Higgs factory~\cite{1402.5973}.
      \end{enumerate}
     
\item $\mu^+\mu^-$ colliders
     \begin{enumerate}
    \item Muon collider - with colliding $\mu^+\mu^-$ beams, already a circular tunnel of $300 \units{m}$  could host a Higgs factory, with the Higgs produced directly in the $s$-channel. The centre of mass energy could be ugraded to $5 \units{TeV}$~\cite{mu-collider}.
         \end{enumerate}
      
 \item $\gamma\gamma$ - these would be possible derivatives of $ee$ colliders. 
 
 \item $ep$ and $eA$ facilities
 
 A compilation~\cite{cerncourrier} of existing and planned lepton-nucleon scattering facilities is presented in Fig.~\ref{fig:eAfacilities}  as a function of center of mass energy and luminosity. 
 
 With the exception of the CERN potential facilities, LHeC~\cite{LHeC} (which requires the addition of an $60 \units{GeV}$ energy recovery linac) and FCCep, the remaining new facilities are to be electron-ion colliders (EIC) with polarized beams. The planned facilities in China, shown in the figure as CEIC1 and CEIC2, belong to the High Intensity Heavy Ion Accelerator project (HIAF) at the Institute of Modern Physics in Lanzhou. The EIC~\cite{EIC-whitepaper} in the US has still to be decided. In the mean time two projects are being developed, the JLab EIC version~\cite{EIC-JLab} (MEIC) and the Brookhaven one~\cite{EIC-RHIC} (eRHIC). At this point the most realistic is the US EIC as it was endorsed by the Nuclear Science Advisory Committee as the next highest priority project for the nuclear physics community in the US.  Compared to HERA, the center of mass energies are lower by a factor two to three, however the luminosity is expected to be higher by three to four orders of magnitude.

 \end{itemize}
 
 The time scale of all these projects is difficult to assess, however it is hard to believe that any new facility will come into existence well before 2030. One needs also to take into account that the choice will be ultimately driven by globalization, consensus and cost.

\section{Future program}
 
\begin{figure}[h]
\centering
\includegraphics[width = 0.7\linewidth]{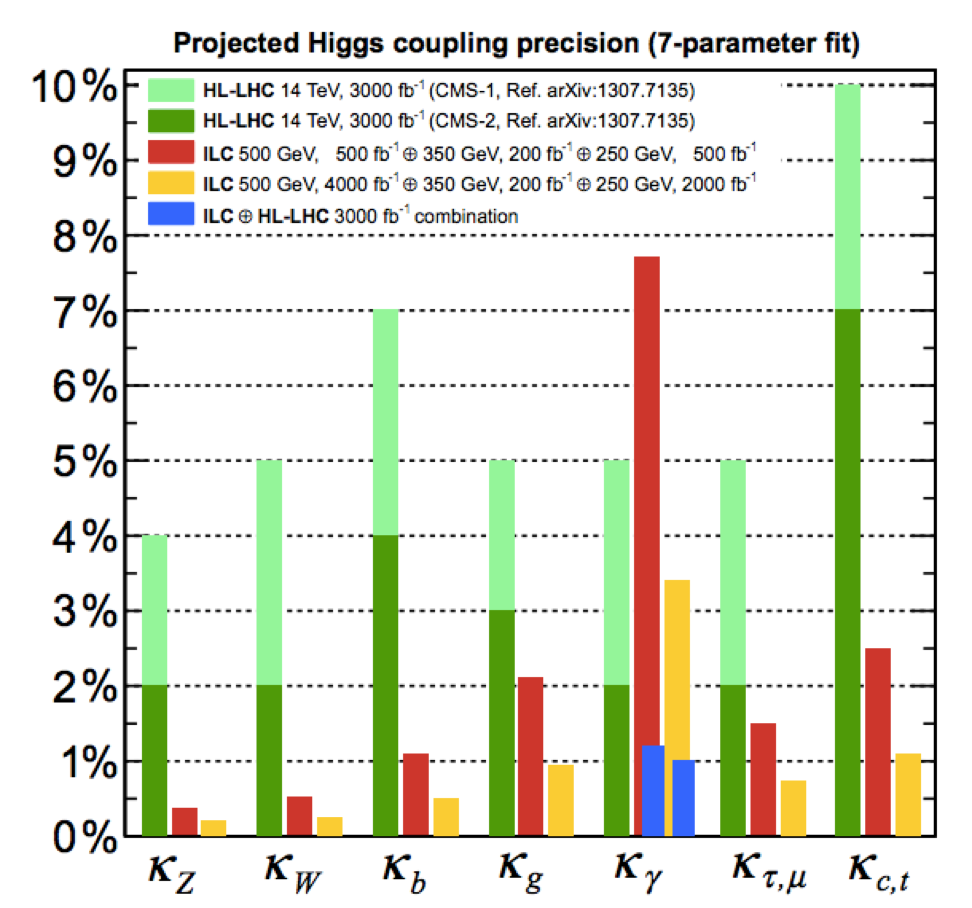}
\caption{Expected precision on Higgs couplings for the ILC compared to the achievable precision at the LHC after luminosity upgrade.}
 \label{fig:Higgscouplings}
\end{figure}

 The mission of the particle physics community is to look for deviations from the Standard Model which could explain dark matter, neutrino masses and the matter/anti-matter asymmetry. One place to look for guidance is precision measurements of the Higgs sector. Here the HL-LHC has a strong competition from the linear colliders~\cite{ILC-physics,CLIC-physics}. This is illustrated in Fig.~\ref{fig:Higgscouplings} where the expected precision of Higgs coupling constants from HL-LHC
 are compared to the ones expected from the ILC, and these in turn are compared to the combined sensistivity. With the exception of the coupling of the Higgs to photons, the ILC will be able to achieve a sensitivity below $1\%$ while at the LHC a precision of $2$ to $3 \%$ is expected.   
 
The  mission of the nuclear physics community is to understand confinement and the generation of mass in QCD as $95\%$ of the nucleon mass is due to strong interactions. The big advantage of the electron-ion colliders is also the fact that they provide vital constraints to the proton structure which is essential for calculating the SM expectations for hadron machines. Therefore their existence is important to the particle physics community as well.
 
In the best of worlds, one would like to have one energy frontier machine (HE-LHC, FCChh, SppC), one precision-frontier machine (ILC, CLIC, CepC, FCCee) and one machine from density and scale frontier (EIC, LHeC, FCCeh). Even better if these machines could be realized on a similar time scale so that their complementarity can be fully exploited.
 
 All these considerations will be part of the European Strategy Update and it is important that all proponents take part in these discussions.
 
 Also part of these considerations should be the future developments in the accelerator technologies because the sustainability of the presently planned accelerators is a big issue.
 
\section{Summary}

We are all hoping that in the near future LHC will bring new discoveries which will guide the community towards future facilities.  Taking into account the extraordinary performance of the LHC accelerator and of the LHC detectors, it is clear that the LHC community will not miss a spectacular discovery. However, the window for large effects is slowly closing. That is where precision measurements of the Standard Model become a real alternative. Even in that case, in spite of tremendous efforts of the theoretical community, the precision that can be achieved at the LHC in some regions of phase space, starting from Higgs production, is limited by the knowledge of the proton structure. It is therefore very much in the interest of us all to strongly support the EIC project in the US. 
 At the same time, we have to be prepared to make a strong physics case for the $e^+e^-$ linear collider independently of whether LHC points the way to new physics. If we want as a community to remain dynamic and attractive to the next generation of explorers, we need to unite and then our perspective is limitless.

\end{document}